\documentclass[10pt,aps,prl,twocolumn,showpacs,superscriptaddress,amsmath,amssymb]{revtex4-1}
\usepackage{bm}
\usepackage{color}
\usepackage{graphicx}
\usepackage{xcolor}

\usepackage{fancyvrb}
\VerbatimFootnotes 

\begin{document}
\title{Robust phase retrieval with the swept approximate message passing (prSAMP) algorithm}

\author{Boshra Rajaei$^{1,5,7}$,
				Sylvain Gigan$^{2,4,5}$, 
				Florent Krzakala$^{3,4,5}$,
                Laurent Daudet$^{1,5,6}$}
\noaffiliation
\affiliation{
 Institut Langevin, ESPCI and CNRS UMR 7587, Paris 75005, France\\
$^{2}$ Laboratoire Kastler Brossel,  CNRS UMR 8552 \&
  \'Ecole Normale Sup\'erieure, Paris 75005, France. \\
$^{3}$ Laboratoire de Physique Statistique, CNRS UMR 8550 \&
  \'Ecole Normale Sup\'erieure, Paris 75005, France.\\
$^{4}$ Sorbonne Universit\'es, UPMC Univ. Paris
06,  Paris 75005, France\\
$^{5}$ PSL Research University, Paris 75005, France\\
$^{6}$ Paris Diderot University, Sorbonne Paris Cit\' e, Paris 75013 , France\\
$^{7}$ Sadjad University of Technology, Mashhad, Iran                    
                   }

\begin{abstract}
In phase retrieval, the goal is to recover a complex signal from the magnitude of its linear measurements. While many well-known algorithms guarantee deterministic recovery of the unknown signal using i.i.d. random measurement matrices, they suffer serious convergence issues some ill-conditioned matrices. 
As an example, this happens in optical imagers using binary intensity-only spatial light modulators to shape the input wavefront. The problem of ill-conditioned measurement matrices has also been a topic of interest for compressed sensing researchers during the past decade. 

In this paper, using recent advances in generic compressed sensing, we propose a new phase retrieval algorithm that well-adopts for both Gaussian i.i.d. and binary matrices using both sparse and dense input signals. This algorithm is also robust to the strong noise levels found in some imaging applications. 

\end{abstract}
\maketitle

\section{Introduction}\label{sec:introduction}
This paper considers the fundamental problem of recovering a complex signal, $\mathbf{x}$, from magnitude of its linear projections. This problem is called \textit{phase retrieval} (PR). Indeed, in many  imaging setups,  detectors (for instance, CCD cameras)  are fundamentally intensity-only. Getting access to phase measurements may not be possible, or may involve a significantly more complex physical setup, e.g. with interferometry. 
 Some of these applications include X-ray crystallography \cite{harrison93}, X-ray diffraction imaging \cite{bunk07}, optical imagers \cite{walther63,liutkus14} and astronomical imaging \cite{fienup87}. PR problems in the presence of additive noise be may formulated  \cite{dremeau15} as:  
\begin{equation}\label{equ:pr}
\mathbf{y}=|\mathbf{H}\mathbf{x}+\mathbf{w}|^2
\end{equation}
where $\mathbf{y}\in \mathbb{R}_+^M$ known (measured) output, $\mathbf{H}$ is the $M \times N$ known complex projection matrix, $\mathbf{x} \in \mathbb{C}^M$ is the unknown input, and 
$\mathbf{w} \in \mathbb{C}^M$ is the "noise" term - upon which some statistical assmuptions are made. 
Many methods have been reported in phase recovery while $\mathbb{H}\in \mathbb{C}^{M\times N}$ is the Fourier 
transform or a random  matrix with iid Gaussian coefficients. 
These methods include, but are not limited to, convex relaxation algorithms such as
phaseLift \cite{candes13} and phaseCut \cite{waldspurger15}, error reduction algorithms such
as Gerchberg and Saxton \cite{gerchberg72} and Fienup \cite{fienup78} and several variants
of them \cite{marchesini07,netrapalli15} and spectral recovery method \cite{alexeev14}.

Here, we are interested in the more challenging problem of recovering $\mathbf{x}\in \mathbb{C}^{N}$ using a binary projection matrix, $\mathbf{H}\in \{0,1\}^{M\times N}$. This is the situation we face in real imaging applications using binary intensity spatial light modulators (SLM) such as digital micromirror devices (DMD) \cite{dremeau15}.  Using these ill-conditioned matrices, one is often faced with convergence issues with most of the afore-mentioned algorithms. 

Signal recovery using ill-conditioned is also a challenging problem in other signal processing fields. Recently, in compressed sensing there have been attempts to reconstruct a sparse signal using generic matrices \cite{vila14,cakmak14,manoel14}. Based on a Bayesian approach to a well-known compressed sensing approach which is called approximate message passing algorithm (AMP) \cite{maleki10}, in \cite{manoel14} the authors develop an idea which demonstrates good convergence properties over ill-conditioned noisy matrices. The extension is called swept AMP (SwAMP).  

Following the Bayesian method, there are also a few phase retrieval methods such as phase retrieval generalized AMP (prGAMP) \cite{schniter15}. By utilizing a magnitude only output prior, prGAMP reaches near optimal results to the classic PR problem with a smaller number of measurements. Phase retrieval variational Bayes expectation maximization (prVBEM) \cite{dremeau15} is a mean-field variational Bayes phase retrieval technique that was developed  for the task of calibrating the transmission matrix of a strongly scattering material, using binary measurements \cite{dremeau15}. Although prVBEM has both small complexity and robustness to strong noise, its application has only been demonstrated in the context of light focusing \cite{rajaei15}.

In this paper, we mix the idea of SwAMP with phase retrieval strategies, in order to solve \eqref{equ:pr} over binary projection matrices. The algorithm is called prSAMP and is already partially presented in \cite{rajaei15}\footnote{C source code to produce the same results as this paper is accessible at corresponding IPOL web page http://www.ipol.im.}. In the context of compressive imaging though scattering material \cite{liutkus14}, 
we here show that prSAMP can effectively deal with the phase-less recovery problem using intensity-only SLM. This  yields to these two different problems in calibration and recovery steps: 1) complex input and binary measurement matrix and 2) binary input and complex measurement matrix. Obviously, the special case of complex input and matrix as addressed by most PR algorithms is also solvable by the algorithm.

\section{Notation}\label{sec:nota}
In this section a brief summary of the notations that is used throughout the paper, is provided. As usual, scalars, vectors and matrices are written in small regular-face, small bold-face and capital bold-face letters, respectively. The $i$th entry of a vector $\mathbf{x}$ is denoted by $x[i]$ and the $i$th column of matrix $\mathbf{H}$ by $\mathbf{h}[i]$. The $\times$ and $\odot$  operators stand for vector and element-wise multiplication. We also use $(.)^{\circ 2}$ and $\oslash$ for element-wise square power and division. In algorithms, a function $p$ is represented as $@p$ that is defined either in text or in another algorithm.   

\section{prSAMP Algorithm}\label{sec:algo}
\begin{figure*}[t]
  \centering
    \includegraphics[width=0.6\textwidth]{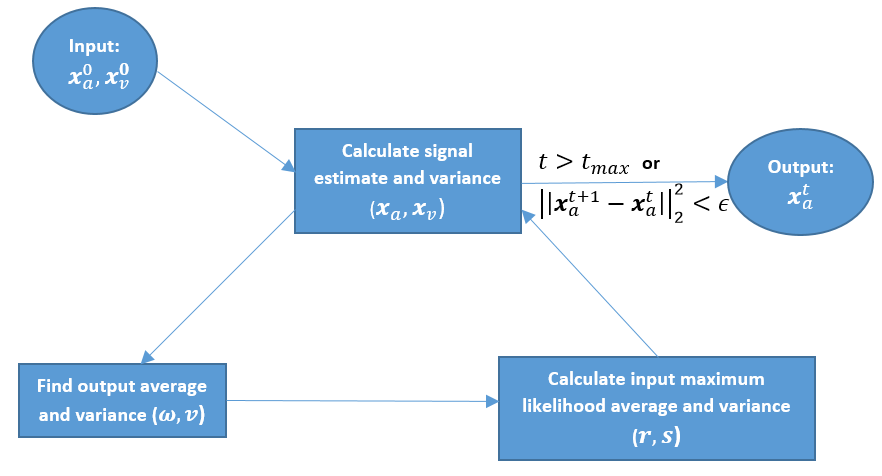}
\caption{Approximate message passing algorithm using a Bayesian statistical approach.}
\label{fig:amp}  
\end{figure*}

The proposed phase retrieval swept approximate message passing (prSAMP) algorithm is a mixture of two ideas in compressed sensing (CS) and phase retrieval, in addition to some modifications to work in 2D image recovery. The first part is swept approximate message passing algorithm (SwAMP) \cite{manoel14}, which is one of many variants of approximate message passing (AMP) method \cite{maleki10} in compressed sensing. In the context of CS, AMP is an iterative algorithm that reconstructs a sparse signal $\mathbf{x}$ from a set of under-determined linear noisy measurements, $\mathbf{y}=\mathbf{H}\mathbf{x}+\mathbf{w}$, where $\mathbf{w}\sim\mathcal{N}(0,\sigma^2)$. Figure \ref{fig:amp} shows the statistical approach in the AMP method \cite{krzakala12}, where the algorithm starts from initial posterior estimates of signal average and variance $\mathbf{x}_a^0$ and $\mathbf{x}_v^0$. It then follows three main steps iteratively; 1) calculate output mean and variance variables, $\bm{\omega}$ and $\mathbf{v}$, 2) calculate input maximum likelihood terms, $\mathbf{r}$ and $\mathbf{s}$, which are also called AMP Gaussian fields and 3) use AMP denoisers to update input signal mean and variance, $\mathbf{x}_a$ and $\mathbf{x}_v$. The AMP denoisers carry prior knowledge of  the input unknown signal. Later in this paper, we define two denoisers for binary and Gaussian signals. AMP has been shown to converge to optimal solution while working with i.i.d. gaussian projection matrices \cite{bayati11}. However, it does not necessarily converge for generic matrices \cite{caltagirone14}. This is where the idea of SwAMP brings up. 

SwAMP is a simple change in step 2 of AMP. Instead of standard parallel calculation, a sequential, or \textit{swept}, random update of AMP maximum likelihood variables is suggested which shows significant stabilization of the AMP loop while working with different non i.i.d. and/or ill-conditioned projection matrices \cite{manoel14}.      

To extend AMP framework to our PR problem of eqn. \eqref{equ:pr}, we first use the generalized AMP (GAMP) \cite{rangan11} which is an extension of AMP for arbitrary output channels, i.e. $\mathbf{y}=q(\mathbf{H}\mathbf{x}+\mathbf{w})$. This adds an output function, $@p_{out}$, which is dependent on the stochastic description of $q(.)$. Normal CS problems follow an additive white Gaussian noise (AWGN) channel as output prior. For the PR case, we follow what is proposed in a GAMP-based phase retrieval algorithm, which is called prGAMP \cite{schniter15} and formulates $@p_{out}$ for $q(|z|)=|z|$ and $q(|z|)=|z|+w$. 

\begin{figure*}[t]
  \centering
    \includegraphics[width=1.0\textwidth]{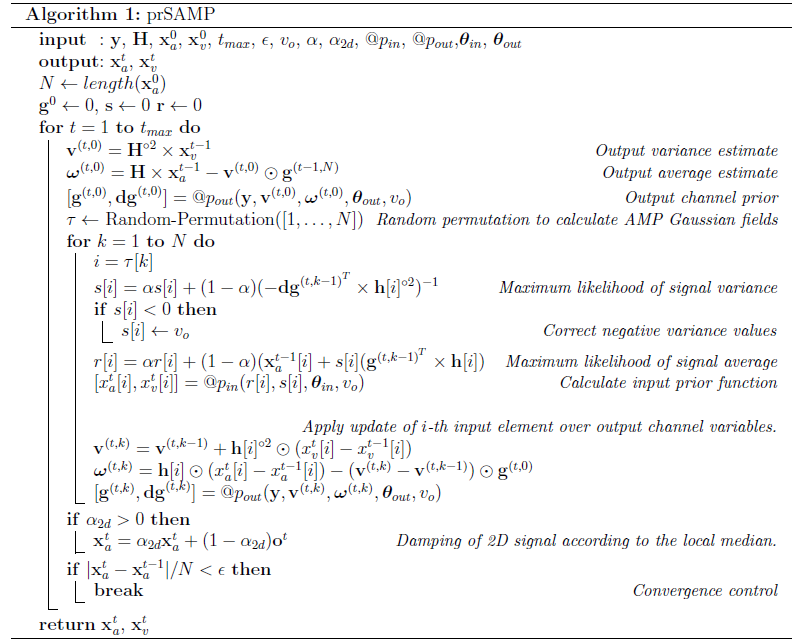}
\label{alg:prSAMP}  
\end{figure*}

Algorithm \ref{alg:prSAMP} describes phase retrieval version of SwAMP, denoted as prSAMP, which combines the swept update ordering and the phase retrieval output channel in the AMP iteration. Beside the intensity measurements, $\mathbf{y}$, and the projection matrix, $\mathbf{H}$, the algorithm has a few other input parameters. These include the two stopping parameters, maximum number of iterations, $t_{max}$, and the precision threshold, $\epsilon$. The algorithm stops if it reaches $t_{max}$ iterations or if the difference between two successive estimations is less than $\epsilon$, $\|\mathbf{x}_a^{t}-\mathbf{x}_a^{t-1}\|_2^2<\epsilon$. The other parameter is $v_0$. During prSAMP iterations, variance terms may become negative or very small. This prevents the algorithm to improve its current estimation which happens often during the first iterations. In these cases the bad variance values replace by $v_0$. There are also two damping parameters, $\alpha$ and $\alpha_{2d}$. Damping is necessary in case of ill-conditioned matrices. We use the first damp factor for $\mathbf{s}$ and $\mathbf{r}$ variables in step 2 and the second for 2D signals. If the input signal is actually vectorized version of a 2D image, after step 3 we add one step to take into account the 2D relation between $\mathbf{x}_a$ elements. Here, we employ a simple damping with respect to $\mathbf{o}^t$ which is local median in iteration $t$ but more sophisticated 2D priors may establish for better smoothness in recovered signal. Finally, we have input and output prior functions and their associated parameters which are defined in separate algorithms. 

\begin{figure*}[t]
  \centering
    \includegraphics[width=1.0\textwidth]{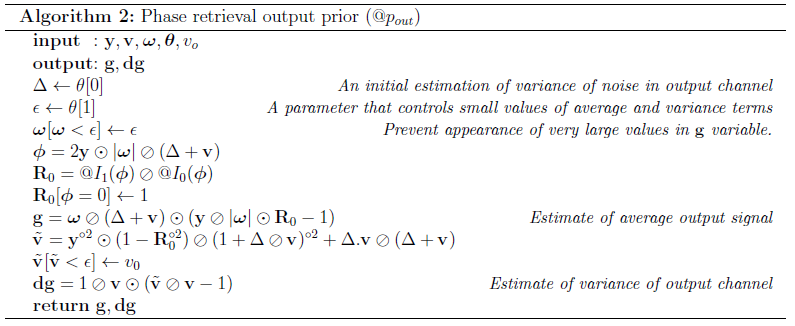}
\label{alg:pout}  
\end{figure*}

In the main loop, the algorithms starts by estimating output average and variance terms, $\bm{\omega}^t$ and $\mathbf{v}^t$. Then we have output prior applies over these variables to calculate $\mathbf{g}^t$ and $\mathbf{dg}^t$ mean and variance terms. In the case of phase retrieval experiment, these are defined distinctly in Algorithm \ref{alg:pout} but in normal compressed sensing with additive white Gaussian noise (AWGN) the calculation is straightforward. AWGN output prior indicates $(\mathbf{y}-\bm{\omega})/(\mathbf{v}+\Delta)$ and $-1//(\mathbf{v}+\Delta)$ for $\mathbf{g}$ and $\mathbf{dg}$ variables, respectively. 

In the second step, we have the sequential swept iteration for maximum likelihood terms, $\mathbf{s}$ and $\mathbf{r}$. It has been claimed in the SwAMP original paper that random computation of involved variables result in better convergence therefore we also follow the same method. After each index $i$ is calculated from input signal, the updates should apply over output channel variables. Finally, the estimate of unknown input signal is returned as $\mathbf{x}_a^t$ variable.

Considering a circular Gaussian additive white noise in measurements, $|\mathbf{y}|$ follows a Rician probability density function which is the basis for a PR output channel derivation in prGAMP paper \cite{schniter15}. We also follow the same formulation. Algorithm \ref{alg:pout} explains the PR output prior. Here, $@I_0(.)$ and $@I_1(.)$ functions are respectively $0^{th}$ and $1^{st}$-order modified Bessel functions of first kind.  

\begin{figure*}[t]
  \centering
    \includegraphics[width=1.0\textwidth]{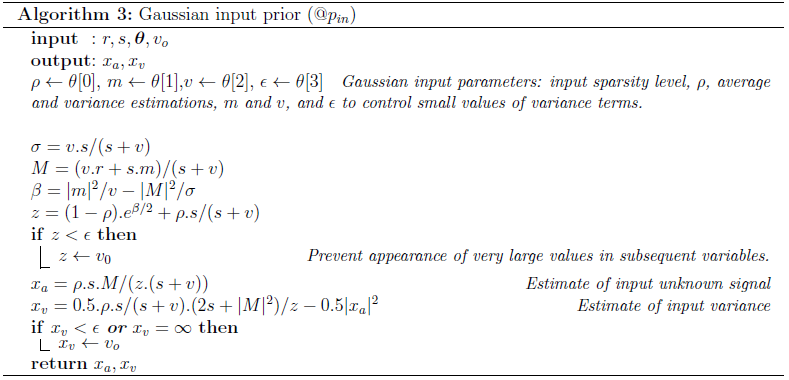}
\label{alg:gpin}  
\end{figure*}

\begin{figure*}[t]
  \centering
    \includegraphics[width=1.0\textwidth]{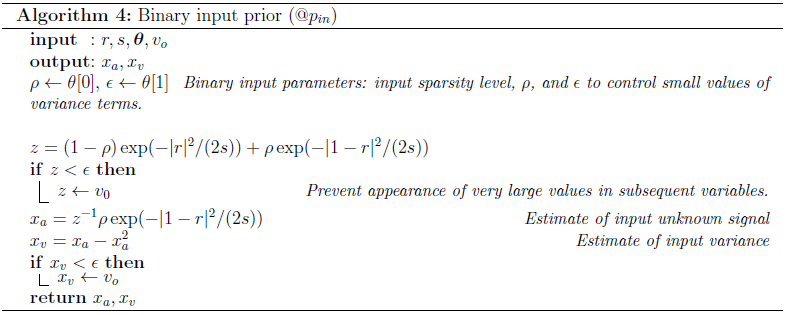}
\label{alg:bpin}  
\end{figure*}

As we mentioned earlier, in this paper we are interested in solving PR problem in two cases: 1) calibration step with $\mathbf{x}\in \mathbb{C}^{N}$ and $\mathbf{H}\in \{0,1\}^{M\times N}$ and 2) recovery step with $\mathbf{x}\in \{0,1\}^{N}$ and $\mathbf{H}\in \mathbb{C}^{M\times N}$. Therefore, a Gaussian input prior for the calibration phase is a reasonable choice as it is described in Algorithm \ref{alg:gpin}. Furthermore, a possible binary prior is explained in Algorithm \ref{alg:bpin} for the reconstruction step.

\section{Implementation}\label{sec:implementation}
To reconstruct the complex signal $\mathbf{x}$ (up to a global phase) using its intensity-only projections, the size $M$ of the measurement vector should be at least $2N$ - it has been established recently that, in a generic case, $M \leq 4N$ measurements are required \cite{bodmann15} to recover a unique $x$. This means that prSAMP follows a computational complexity of $O(N^3)$ which is a bottleneck for real-time imaging. Due to sequential nature of swept loop we can not solve this scaling issue directly but there are two possibilities to alleviate it: 1) in calibration phase since different rows of projection matrix are inherently independent, the algorithm is fully parallel. In the supplementary files, two extensions of prSAMP using OMP and MPI parallel tools, are provided. 2) the other enhancement option is an idea we call block-based phase retrieval \cite{rajaei16}. This block-based PR method starts by splitting the $M\times N$ input  problem into $K$, $m_i\times n_i$ sub-problems, where $\sum_{i=0}^{K-1}n_i=N$, $m_i=\lceil \alpha n_i \rceil$ and $\alpha=M/N$. The $K$ sub-problems are then solved in parallel. Finally, all the partial results are merged with a few extra global measurements, by applying a low-dimension global phase tuning step. In this way the order of prSAMP algorithm breaks down into $O(N^3/K^2)$. This comes at a price of being able to design the measurement matrix in a general block-diagonal manner which is the case in any physical systems
where one can probe the whole object by parts. Block-based prSAMP is extended in the supplementary files using Matlab.   

\section{Parameter study}\label{sec:param}
There are two groups of parameters: first, the main prSAMP parameters and second, priors parameters. Depending on prior knowledge of input and output signals, $\mathbf{x}$ and $\mathbf{y}$, we may require to provide parameters like, noise variance in measurements, $\Delta$, an estimation of input sparsity level, $\rho$, or input mean and variance, $m$ and $\sigma$, in case of Gaussian input prior. Beside these obvious prior-dependent parameters, there are a few main parameters that they play important role in algorithm convergence.  The initial estimation of unknown input signal, $\mathbf{x}^0$, and the damping factor, $\alpha$, are the two most important ones. 

As it is well-known the compressive phase retrieval problem generally suffers from convergence to  local minima \cite{vila14}.  Empirical studies show that the situation is worse while working with ill-conditioned non-gaussian i.i.d. random matrices \cite{manoel14}. In case of Gaussian input signals, like what we have in the calibration phase, using a pseudo random generator to initialize $\mathbf{x}_a^0$ seems a reasonable choice. Afterwards if the algorithm diverges, a complete restart with a new random initial vector is necessary. Multiple restarts was first suggested in prGAMP paper \cite{schniter15}. The solution that yields to lowest normalized residual, $NR=\|\mathbf{y}-|\mathbf{A\hat{x}}|\|_2^2/\|\mathbf{y}\|_2^2$, is selected as the algorithm output.  In case of other types of input signals, a good initial point would guarantee the algorithm convergence. For example, in recovery phase of our optical imager, we employ a low resolution (LR) version of input image. This LR signal may be gathered from negative output of DMD array or numerically estimated based on specific image database. For signals of length $N=2^{14}$, LR version of $2^6$ is a good start point. Depending on initial point confidence, $\mathbf{x}_v^0$ variance vector is selected from $(0,1]$ interval. In calibration and recovery steps we empirically set $\mathbf{x}_v^0$ values at 0.5 and 0.1, respectively. 

The other important parameter is the damping factor, $\alpha$. Damping slows down the convergence rate of the algorithm, and hence prevents being stuck into a possibly wrong local minima,  while still keeping information from previous iterations. Here, $\alpha$ is a scalar from $[0,1)$ interval where 0 indicates no damping situation. In case of ill-conditioned projection matrices we need more damping. In our experiments we use 0.9 and 0.2 values for calibration and recovery steps, respectively. 

In case of 2D input signals we have another damping parameter, $\alpha_{2d}$. In this paper as a 2D prior we used a simple damping step which mixes the current solution,$x[i]_a^t$, with a representative of its neighborhood, $o[i]^t$. The representative is median over a $5 \times 5$ block centered at element $i$. This may improve by taking into account learned priors like the RBM prior as it is proposed in \cite{tramel15}. The more sophisticated priors usually come with the price of an offline learning step. Hence, since our simple damping prior provides satisfactory results, as it is shown in the next section, we left further improvements to the interested reader.

The other parameters include: maximum number of iterations, $t_{max}$, precision factor, $\epsilon$, and negative variance factor, $v_0$. Number of iterations is usually a factor of number of nonzero elements in input unknown signal, $\rho N$. In calibration step, with a full rank input vector, we set $t_{max}$ at $N/4$ empirically. But for small $N$, it is necessary to let algorithm pass the initial oscillations. In small $N$, we may use $t_{max} \leftarrow N$. 

Precision factor, $\epsilon$, is another measure of convergence which ensures a minimum difference between two successive solutions. A difference less than $\epsilon$ indicates the algorithm is iterating around a local minima and, Hence, there is no progress. 

Finally, negative variance factor, $v_0$, is employed in case of resulting a negative variance term. There are various variance variables in prSAMP algorithm like: $s$ and $\omega$ variables in the main algorithm and $\tilde{v}$, $z$ and $x_v$ variables in priors. These terms have to be positive and not extremely small. Therefore, in case of negative or very small variance terms, $v_0$ is used as a replacement value. This parameter should be sufficiently large and in the range of $x_v^0$ because negative variance indicates a bad situation in prSAMP iteration and we should set the variance at a large value to let algorithm converge to another mean point.  

\begin{figure*}[t]
  \centering
    \includegraphics[width=1\textwidth]{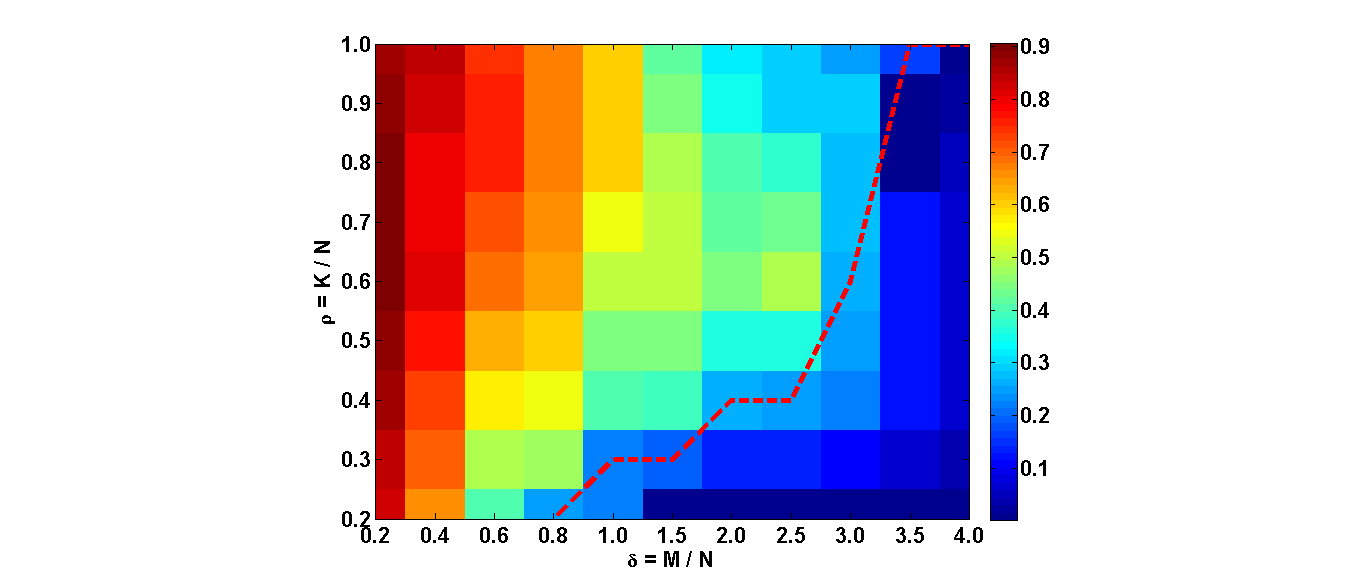}
\caption{prSAMP phase transition plot for solving the phase retrieval problem using a binary measurement matrix. Here, $N=256$ and an SNR of 30 dB is considered in all experiments. The performance criterion is NMSE which is selected out of 50 independent trials. The red dashed line represents a transition from failure to success by applying a threshold of 0.2.}
\label{fig:PT1}  
\end{figure*}   

\begin{figure*}[t]
  \centering
    \includegraphics[width=1\textwidth]{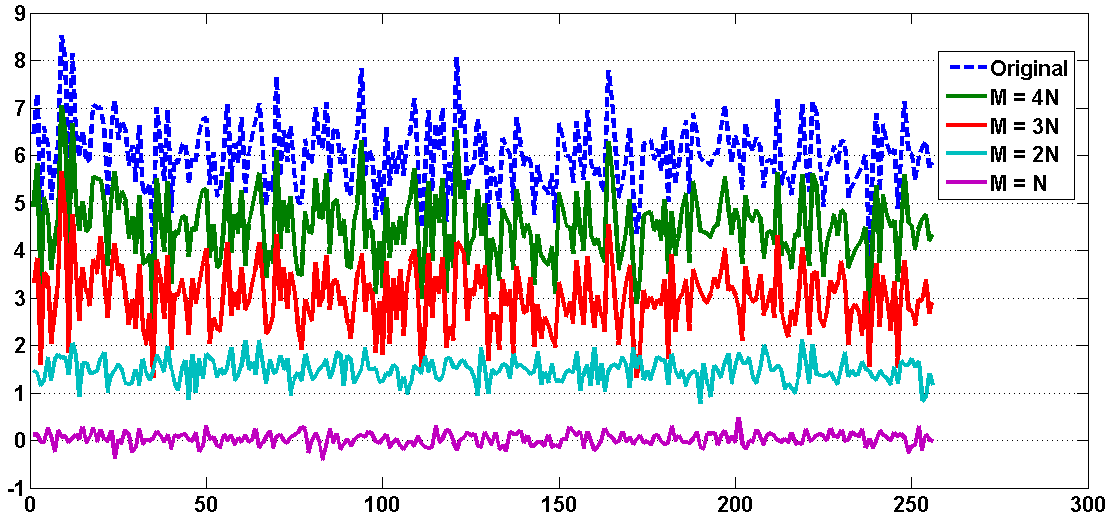}
\caption{An instance signal $\mathbf{x}\in \mathbb{C}^N$ ($\rho=1$) and its four prSAMP reconstructions at $\delta=\{1, 2, 3 $ and $4\}$ using a binary measurement matrix (different offsets are applied for presentation purposes). The real part is plotted and the imaginary part has similar behavior. Complete recovery happens at $\delta=4$.}
\label{fig:complex}  
\end{figure*} 

\begin{figure*}[t]
  \centering
    \includegraphics[width=1\textwidth]{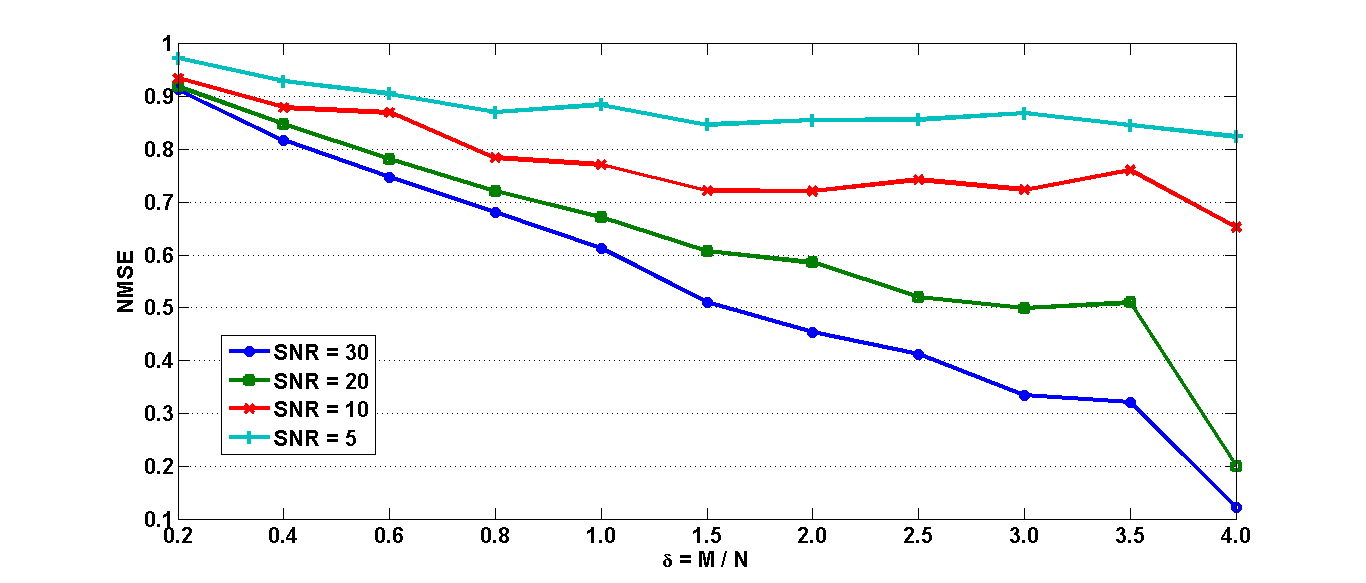}
\caption{The effect of noise on prSAMP performance, as a function of the measurement sampling factor $\delta = M/ N$,  using a binary measurement matrix. 
}
\label{fig:snr}  
\end{figure*} 

\section{Experimental results}\label{sec:exprim}
In this section we investigate the application of using prSAMP algorithm to solve phase retrieval problem \eqref{equ:pr} in two different situations; first, $\mathbf{H}\in \{0,1\}^{M\times N}$ and $\mathbf{x}\in \mathbb{C}^N$ and second, $\mathbf{H}\in \mathbb{C}^{M\times N}$ and $\mathbf{x}\in \{0,1\}^N$. Using binary transition matrix to recover complex input signal, figure \ref{fig:PT1} shows the phase transition plot for $N=256$ and snr equal to 30 dB ($\Delta=10^{-3}$). The error is measured in terms of normalized mean square error (NMSE) between original and the recovered signal after compensating the global phase shift. Each point in the plot is the lowest NMSE obtained by prSAMP in 50 distinct trials. As a result of ill-conditioned binary measurement matrix, the damping factor, $\alpha$ is set to 0.9. A phase transition curve is generated by applying a NMSE threshold of 0.2. The plot confirms the recently established rate of $M \geq 4N$ to reconstruct dense signal $\mathbf{x}$ in phase retrieval regime. The effect of increasing the number of measurements on recovered signal is shown in figure \ref{fig:complex} using the same settings as the previous experiment and a random input. Here, we have $K=N$.    

For $\rho=1$, reconstruction performance of prSAMP is studied at four different noise levels (snr equal to 30, 20, 10 and 5 dB) and $0.2\leq \delta \leq 4.0$ in figure \ref{fig:snr}. According to this experiment in case of strong noisy measurements, after $\delta=1$ adding more samples does not improve the results significantly.

\begin{figure*}[t]
  \centering
    \includegraphics[width=1\textwidth]{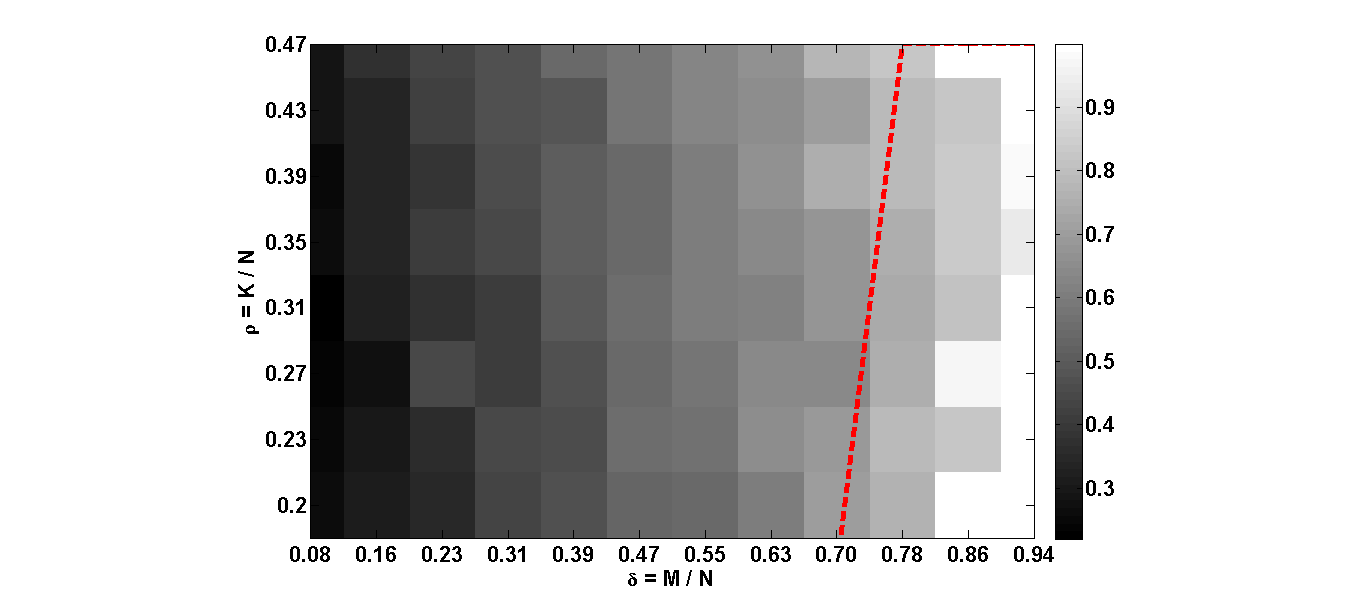}
\caption{prSAMP phase transition plot for solving the phase retrieval problem, with a binary input, using a complex measurement matrix. Here, $N=256$ and an SNR of 30 dB is considered in all experiments. The performance criterion is the correlation, which is selected best out of 50 independent trials. The red dashed line represents a transition from failure to success by applying a threshold of 0.8.}
\label{fig:PT2}  
\end{figure*} 

\begin{figure*}[t]
  \centering
    \includegraphics[width=0.8\textwidth]{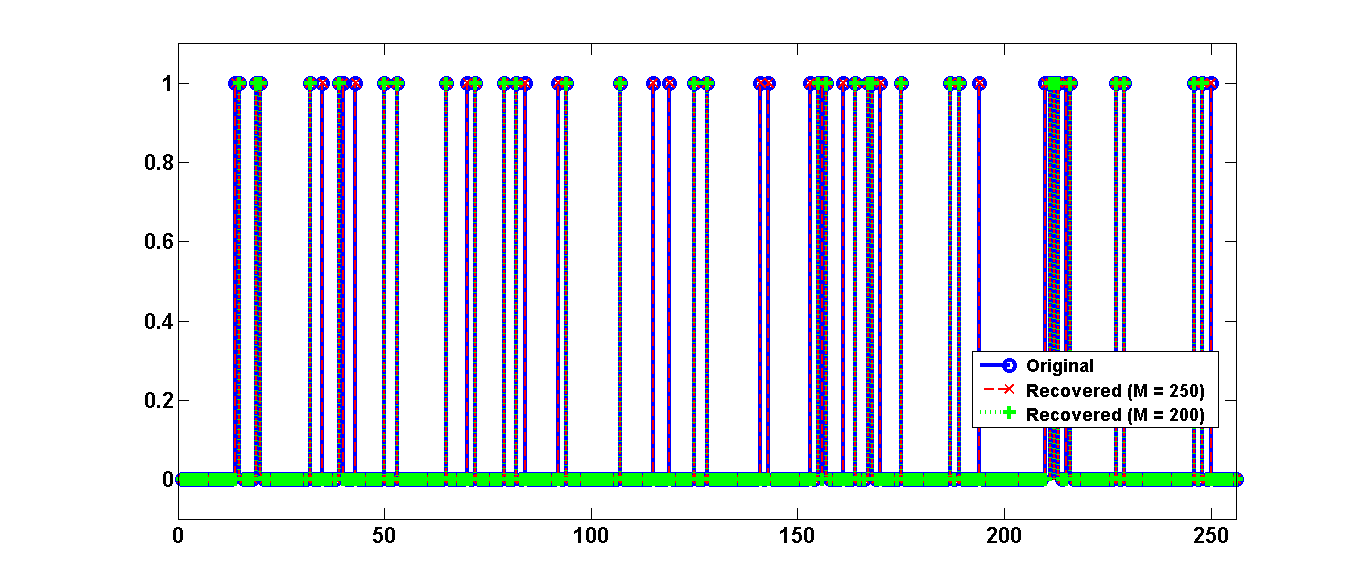}
\caption{An instance signal $\mathbf{x}\in \{0,1\}^N$ ($K=50$) and its two prSAMP reconstructions at $M=\{200 $ and $250\}$ after applying a threshold of 0.5.}
\label{fig:binary}  
\end{figure*} 

In the second experiment, prSAMP is applied to the problem of reconstructing binary random input signals. Figure \ref{fig:PT2} shows the corresponding phase transition plot. Except $\alpha$ which is set to 0.2 the other parameters are similar to the first experiment. The recovery error is measured in terms of best correlation out of 50 runs. Here, the number of necessary measurements for complete recovery is decreased significantly at different sparsity levels probably due to binary input prior. This fact also has been shown in figure \ref{fig:binary} which plots a random binary signal with $K=50$ and its two reconstructions at $M=200$ and $M=250$. Comparing to figure \ref{fig:complex} complete recovery is happened using significantly less measurements ($M=N$). 

\section{Computational complexity}
Finally, it would be interesting to have a brief discussion on computational complexity of prSAMP algorithm. Even though, as our experiments show, the algorithm performs well for ill-conditioned matrices and strong noise situations, it does not scale well as the size of input signal increases. In Algorithm \ref{alg:prSAMP} the number of iterations, $t_{max}$, and measurements, $M$, grow linearly with input size $N$. Therefore, prSAMP follows a cubic $O(N^3)$ computational complexity. In addition to this, the amount of data that the algorithm has to handle at least scales with $O(N^2)$. This is challenging at large inputs. In \cite{rajaei16}, a block-based version of prSAMP has been proposed that can reduce computational cost and also memory requirements of original prSAMP by several orders of magnitude.

\section{Conclusion}
In this study, a new phase retrieval algorithm has been proposed, called phase retrieval swept AMP (prSAMP). prSWAMP is here  numerically evaluated in two situations inspired by real imaging setups. 
In particular, prSWAMP solves the challenging problem of estimating a complex input signal using binary patterns. In reverse, we also show that prSAMP accurately estimates a binary unknown signal using a complex transmission matrix.  

\section*{Acknowledgment}
This research has received funding from PSL Research University under contract CSI:PSL, from the European Research Council under the EU'€™s 7th Framework Programme (FP/2007- 2013/ERC Grant Agreement 
307087-SPARCS and 278025- COMEDIA) ; and from LABEX WIFI under references 
ANR-10-LABX-24 and ANR-10-IDEX-0001-02-PSL$^\star$.

This work was also granted access to the HPC resources of MesoPSL financed by the Region Ile de France and the project Equip@Meso (reference ANR-10-EQPX-29-01) of the programme Investissements d'Avenir supervised by the Agence Nationale pour la Recherche.


\bibliographystyle{siam}
\bibliography{references}

\end{document}